\documentclass[aps, pra, a4paper, preprint, superscriptaddress, showpacs]{revtex4-1}

\usepackage{amsmath,amssymb}
\usepackage{bm}
\usepackage{graphicx}
\usepackage[dvips]{color}

\begin{document}

\title{Theoretical investigations on magnetic field induced $2p^53s~^3P_{0,2} - 2p^6~^1S_0$ transitions in Ne-like ions without nuclear spin}

\author{Jiguang Li}
\affiliation{Department of Physics, Lund University, S-221 00 Lund, Sweden}
\author{Jon Grumer}
\affiliation{Department of Physics, Lund University, S-221 00 Lund, Sweden}
\author{Wenxian Li}
\affiliation{Institute of Modern Physics, Fudan University, 200433 Shanghai, China}
\author{Martin Andersson}
\affiliation{Institute of Modern Physics, Fudan University, 200433 Shanghai, China}
\author{Tomas Brage}
\affiliation{Department of Physics, Lund University, S-221 00 Lund, Sweden}
\author{Roger Hutton}
\affiliation{Institute of Modern Physics, Fudan University, 200433 Shanghai, China}
\author{Per J\"onsson}
\affiliation{Group for Materials Science and Applied Mathematics, Malm\"o University, S-205 06 Malm\"o, Sweden}
\author{Yang Yang}
\email{yangyang@fudan.edu.cn}
\affiliation{Institute of Modern Physics, Fudan University, 200433 Shanghai, China}
\author{Yaming Zou}
\affiliation{Institute of Modern Physics, Fudan University, 200433 Shanghai, China}

\date{\today}

\begin{abstract}
We report theoretical results for magnetic field induced $2p^53s~^3P_{0,2} ~-~ 2p^6~^1S_0$ E1 transitions in Ne-like ions with zero nuclear spin ($I=0$) between Mg III and Zn XXI as well as in Ne I. We demonstrate that it is important to include both ``perturber'' states $2p^53s~^1P_1$ and $2p^53s~^3P_1$ in order to produce reliable transition rates. Furthermore, we investigate the trends of the rates along the isoelectronic sequence of the $2p^53s~^3P_{0,2} ~-~ 2p^6~^1S_0$ transitions and their competition with the $2p^53s~^3P_0 ~-~ 2p^53s~^3P_1$ M1 and the $2p^53s~^3P_2 ~-~ 2p^6~^1S_0$ M2 decays. For the $2p^53s~^3P_0$ state the magnetic field induced transition becomes the dominant decay channel for the light elements even in a relatively weak magnetic field, and it will therefore prove useful in diagnostics of the strength of magnetic fields in different plasmas. The influence of an external magnetic field on the lifetime of the $2p^53s~^3P_2$ state is much smaller but still observable for the ions near the neutral end of the sequence. As a special case, the magnetic field effect on the lifetimes of $2p^53s~^3P_{0,2}$ states of neutral $^{20}$Ne is discussed. It is found that the lifetimes are drastically reduced by a magnetic field, which may be an underlying reason for the discrepancies in the lifetime of the $2p^53s~^3P_2$ state between experiment (14.73(14) s) and theory (17.63 s).
\end{abstract}


\maketitle

\section{Introduction}
The effects of magnetic fields are important in many astrophysical or laboratory plasmas and their strengths are crucial plasma parameters~\cite{Johns-Krull1999, Adams2002, Stambulchik2007}. It is well-known that the interaction between the magnetic field and an atom (or ion) causes spectral lines to split into groups of lines (Zeeman splitting), which can be used to determine the magnetic field strength in a plasma~\cite{Adams2002, Stambulchik2007}. On the other hand, the magnetic interaction also breaks the symmetry of an atomic system allowing atomic states with the same magnetic quantum number and parity to mix and bring about ``unexpected" lines to appear in spectra~\cite{Andrew1967, Wood1968} and lifetimes of long lived state to be shortened~\cite{Feldman1967, Levitt1971}. We will refer to these as magnetic-field induced transitions (MITs). In 2003, Beiersdorfer \textit{et al.} identified a magnetic-field induced transition in Ne-like Ar using the EBIT-II electron beam ion trap in the Lawrence Livermore National Laboratory~\cite{Beiersdorfer2003}. They illustrated that the MIT can also be used as a diagnostic of magnetic field strength for high-temperature plasmas.

\begin{figure}[!ht]
\includegraphics{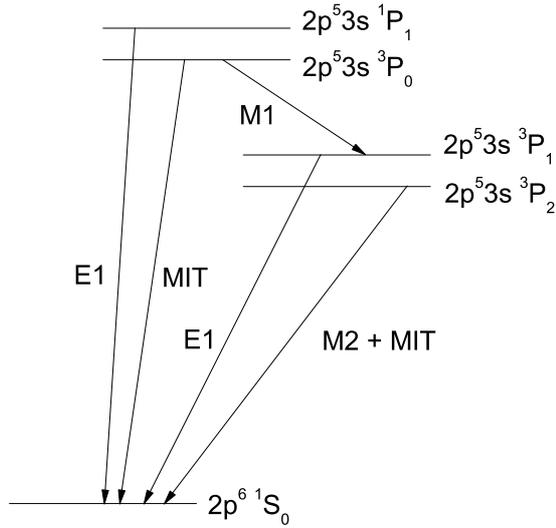}
\caption{\label{EL} The level structure and transitions for the five lowest states of a Ne-like system.}
\end{figure}

Considering the significance of the determination of the magnetic field strengths in plasmas and the promising diagnostic method using MIT lines, we initiated a project to systematically calculate the rates of magnetic field induced $2p^53s~^3P_{0,2} ~-~ 2p^6~^1S_0$ E1 transitions in Ne-like ions with zero nuclear spin ($I=0$) between Mg III and Zn XXI. The transitions in neutral Ne, as a special case, are investigated as well for the unresolved discrepancies in the lifetime of the $2p^53s~^3P_2$ state~\cite{Li2012}. For ions with nuclear spin the hyperfine interaction also induces transitions from $2p^53s~^3P_{0,2}$ to $2p^6~^1S_0$~\cite{Li2012, Andersson}. However, since the aim of this work is to investigate the MIT's, we do not consider these transitions in the present work. The calculations are performed with GRASP2K~\cite{Jonsson2013} and HFSZEEMAN~\cite{Andersson2008} packages based on the multiconfiguration Dirac-Hartree-Fock (MCDHF) method~\cite{Grant2007}. We emphasize the importance of different perturbers that give rise to the MIT. In addition, the systematic behavior of lines that dominate the decays of the $2p^53s~^3P_0$ and $2p^53s~^3P_2$ levels, namely, magnetic field induced $2p^53s~^3P_{0,2} ~-~ 2p^6~^1S_0$ E1, $2p^53s~^3P_0 ~-~ 2p^53s~^3P_1$ M1 and $2p^53s~^3P_2 ~-~ 2p^6~^1S_0$ M2 transitions (cf. Figure~\ref{EL}) are investigated along the Ne I isoelectronic sequence.

\section{Theoretical methods and computational models}
\subsection{General theory}
In the presence of an external magnetic field ${\bf B}$, the Hamiltonian of an atom without nuclear spin is
\begin{eqnarray}
\label{H}
H = H_{fs} + H_{m},
\end{eqnarray}
where $H_{fs}$ in our approach is the relativistic fine-structure Hamiltonian including the Breit interaction and parts of quantum electrodynamical (QED) effects, and $H_{m}$ is the interaction Hamiltonian with the external magnetic field. If the magnetic field is homogeneous through the atomic system, the interaction Hamiltonian is expressed by~\cite{Cheng1985}
\begin{eqnarray}
H_m = ({\bf N}^{(1)} + \Delta {\bf N}^{(1)}) \cdot {\bf B},
\end{eqnarray}
where the last term is the so-called Schwinger QED correction. For an $N$-electron atom the tensor operators are given by
\begin{eqnarray}
{\bf N}^{(1)} &=& \sum_{k=1}^N {\bf n}^{(1)}(k) = \sum_{k=1}^N -i \frac{\sqrt{2}}{2\alpha} r_k ({\bm \alpha}_k {\bm C}^{(1)}(k))^{(1)}, \\
\Delta{\bf N}^{(1)} &=& \sum_{k=1}^{N} \Delta {\bf n}^{(1)}(k) = \sum_{k=1}^{N} \frac{g_s - 2}{2} \beta_k {\bm \Sigma}_k,
\end{eqnarray}
where $\bm \Sigma_k$ is the relativistic spin-matrix and $g_s = 2.00232$ the $g$-factor of the electron spin corrected for QED effects. In addition, $\alpha$ is the fine-structure constant and ${\bm \alpha}$ and $\beta$ are Dirac matrices.

We choose the direction of the magnetic field as the $z$-direction, making $M$ the only good quantum number. The $M$-dependent atomic state wavefunction $| M \rangle$ can be written as an expansion
\begin{eqnarray}
\label{AW}
|M \rangle = \sum_{\Gamma J} d_{\Gamma J} |\Gamma J M \rangle.
\end{eqnarray}
where the $|\Gamma J M \rangle$ are atomic state wave functions (ASFs) that are eigenstates of the Hamiltonian $H_{fs}$. The coefficients $d_{\Gamma J}$ are obtained either through perturbation theory or by solving the eigenvalue equation using HFSZEEMAN package~\cite{Andersson2008}. In terms of perturbation theory, $d_{\Gamma J}$ is given, to first order, by
\begin{eqnarray}
\label{MC}
d_{\Gamma J} = \frac{\langle \Gamma J M |H_m| \Gamma_0 J_0 M_0 \rangle}{E(\Gamma_0 J_0) - E(\Gamma J)},
\end{eqnarray}
where $|\Gamma_0 J_0 M_0\rangle$ represents the reference atomic state.

For an atomic system in an external magnetic field, electric dipole (E1) transition probabilities from a state $|M' \rangle$ to another state $|M \rangle$ are expressed in s$^{-1}$ by
\begin{equation}
\label{MT}
A = \frac{2.02613 \times 10^{18}}{\lambda^3} \sum_{q} \left |\langle M | P^{(1)}_q | M' \rangle \right |^2.
\end{equation}
Substituting Eq. (\ref{AW}) into Eq. (\ref{MT}), we have
\begin{equation}
\label{MT-2}
A = \frac{2.02613 \times 10^{18}}{\lambda^3} \sum_q \left |\sum_{\Gamma J} \sum_{\Gamma' J'} d_{\Gamma J} d'_{\Gamma' J'}\begin{pmatrix}
                                                  J  & 1 & J'    \\
                                                 -M  & q & M'
                                              \end{pmatrix} \langle \Gamma J || {\bf P}^{(1)} || \Gamma' J' \rangle \right |^2.
\end{equation}
where $\lambda$ is wavelength in \AA, and the square of the reduced matrix element of the electric dipole transition ${\bf P}^{(1)}$ operator is basically the line strength of the corresponding transition (in a.u.)~\cite{Johnson1995}. Since the magnetic interaction mixes atomic states with different $J$ by 1, the selection rule on the total angular momentum for E1 transitions can be extended to $\Delta J = J - J' = 0, \pm 1, \pm 2, \mbox{and} \pm 3$. As an examle, one-photon forbidden E1 $J=0 \leftrightarrow J'=0$ transitions are induced by an external magnetic field.

In the framework of the MCDHF method~\cite{Grant2007}, atomic state wave functions $|\Gamma J M \rangle$ can be expressed by a linear combination of configuration state functions (CSFs) with same parity $P$, total angular momentum $J$ and its $z$-component $M$, that is,
\begin{equation}
|\Gamma J M \rangle = \sum_{i} c_i |\gamma_i J M \rangle,
\end{equation}
where $c_i$ stands for the mixing coefficient, and the label $\Gamma$ is often the same as the label $\gamma_i$ of the dominating CSF. The latter denotes other appropriate quantum numbers of the CSFs such as orbital occupancy and coupling tree. CSFs are built up from products of one-electron relativistic orbitals. Applying the variational principle, one are able to obtain one-electron relativistic orbitals and mixing coefficients from a self-consistent field (SCF) procedure. The Breit interaction
\begin{equation}
\label{BI}
B_{ij} = -\frac{1}{2r_{ij}}\left [{\bm \alpha}_i \cdot {\bm \alpha}_j + \frac{({\bm \alpha}_i \cdot {\bm r}_{ij})({\bm \alpha}_j \cdot {\bm r}_{ij})}{r^2_{ij}} \right ]
\end{equation}
and leading QED effects are taken into account in subsequent relativistic configuration interaction (RCI) computations.

\subsection{MITs in Ne-like ions}
For Ne-like ionic systems in an external magnetic field, the reference states $2p^53s~^3P_{0,2}$ are approximately expressed as
\begin{eqnarray}
\label{WF-1}
&& |``2p^53s~^3P_{0,2}" ~ M  \rangle \nonumber \\
&& = d_0 | 2p^53s~^3P_{0,2} ~ M \rangle + \sum_{\mathcal{S}(=1,3)} d_{\mathcal{S};J=1} |2p^53s~^{\mathcal{S}}P_1 ~ M \rangle.
\end{eqnarray}
The quotation marks on the left-hand side emphasize the fact that the notation is just a label indicating the dominant component of the eigenvector. It should be pointed out that the mixing coefficients $d$ are different between $|``2p^53s~^3P_0" ~ M \rangle$ and $|``2p^53s~^3P_2" ~ M \rangle$ states. Remaining interactions between $2p^53s~^3P_{0,2}$ and other atomic states are neglected due to large energy separations and comparatively weak magnetic interaction couplings. The ground state is very well isolated from other states and it is therefore a good approximation to assume that its wave function can be written as
\begin{equation}\label{WF-2}
|``2p^6~^1S_0" ~M \rangle = |2p^6~^1S_0 M \rangle.
\end{equation}

The mixing with the "perturber" states of $2p^53s~^1P_1$ and $2p^53s~^3P_1$ in Eq. (\ref{WF-1}) opens up the one-photon $2p^53s~^3P_{0,2} ~-~ 2p^6~^1S_0$ E1 transitions. Inserting the angular quantum numbers for the states into Eq. (\ref{MT-2}) and evaluating the 3-$j$ symbol gives the magnetically induced transition rate $A_{MIT}$ as
\begin{eqnarray}
\label{MIT-4}
A_{MIT} &=& \frac{2.02613 \times 10^{18}} {3 \lambda^3} \left |\sum_{\mathcal{S}(=1,3)} d_{\mathcal{S}} \langle 2p^6~^1S_0 || {\bf P}^{(1)} || 2p^53s~^{\mathcal{S}}P_1 \rangle \right |^2.
\end{eqnarray}

For weak to moderate magnetic fields where Eq. (\ref{MC}) from the first-order perturbation theory holds, $d_{\mathcal{S}}$ is proportional to the magnetic field strength, $B$, and we can define a reduced coefficient $d_{\mathcal{S}}^R$ and reduced transition rate, $A_{MIT}^R$, which are independent of $B$ through
\begin{eqnarray}
\label{RMIT}
d_{\mathcal{S}} = d_{\mathcal{S}}^R B,\ \ \ \ A_{MIT} = A_{MIT}^R B^2.
\end{eqnarray}

\subsection{\label{CM} Computational models}
The calculations are carried out by using the same computational strategy as described in Ref.~\cite{Jonsson2011, Jonsson2013a}. The active set method is used to construct the configuration space. Configuration expansions are generated by single (S) and double (D) replacements of orbitals in the reference configurations with ones in an active set. In the present work, a single reference configuration model is adopted as a starting description for the ground and excited states, and the $1s$ core shell is kept closed. The configuration spaces are therefore obtained by SD excitations from the remaining shells of the single reference configurations to the active set. The active set is augmented layer by layer until $n=7$. We impose the restriction on the expansion at the last step ($n=7$) for the excited states that we allow at most one excitation from $2s$ or $2p$. Considering the stability problems in the SCF procedure we optimize only the orbitals in the last added correlation layer at the time (together with mixing coefficients). The RCI computations following the SCF calculations take into account the residual correlations as well as the Breit interaction and the QED corrections, since the configuration spaces are further expanded by including the CSFs obtained from all SD excitations to the $n=7$ orbital set and with triple (T) excitations up to the $n=4$ orbital set. With this model high quality atomic state wave functions based on expansions of several hundred thousand CSFs have been produced, which give excellent transition energies with only 0.011\% errors compared with highly accurate measurements available. To further check the effect of electron correlations, the Breit interaction and the QED effects on the magnetic interaction, we present in Table~\ref{MCA} off-diagonal reduced matrix elements $W = \langle \Gamma J || {\bf N}^{(1)} + {\Delta \bf N}^{(1)}|| \Gamma' J' \rangle$ for $^3P_0$ and $^3P_2$ states in the cases of Mg III and Zn XXI as well as reduced associated mixing coefficients $d_{\mathcal{S}}^R$. It is worth noting that the Breit interaction affects the off-diagonal reduced matrix elements at the low-$Z$ end to some extent. As the atomic number increases, the Breit interaction becomes less relative important and thus the effect reduces. On the other hand, the mixing coefficients are influenced by both electron correlation and Breit interaction effects, since fine structure splittings in the $2p^53s$ configuration are sensitive to these.


\begin{table}
\caption{\label{MCA} Off-diagonal reduced matrix elements $W$ (in a.u.) of the magnetic interaction together with reduced mixing coefficients $d_{\mathcal{S}}^R$ (in T$^{-1}$) for Ne-like Mg and Zn ions. DHF: single configuration Dirac-Hartree-Fock, RCI: relativistic configuration interaction, BI: Breit interaction, QED: quantum electrodynamics effects. Numbers in square brackets are the power of 10.}
\begin{tabular}{ccccccccccccccccccccccccc}
\hline
\hline
   & \multicolumn{2}{c}{($^3P_0,~^3P_1$)} && \multicolumn{2}{c}{($^3P_0,~^1P_1$)} && \multicolumn{2}{c}{($^3P_2,~^3P_1$)} && \multicolumn{2}{c}{($^3P_2,~^1P_1$)} \\
\cline{2-3}\cline{5-6}\cline{8-9}\cline{11-12} Model &  $W$  & $d_{1}^R$ &&  $W$  & $d_{3}^R$ && $W$ & $d_1^R$ && $W$ & $d_3^R$ \\
\hline
\multicolumn{12}{c}{\bf Mg$\bf ^{2+}$} \\
DHF   & $-$0.3972  & $-$3.5252[$-$4] &&  $-$0.09869   & 2.5210[$-$5]  && $-$0.3440 & 2.4997[$-$4] && $-$0.08523 & 1.3281[$-$5] \\
RCI   & $-$0.3972  & $-$3.3507[$-$4] &&  $-$0.09869   & 2.5721[$-$5]  && $-$0.3440 & 2.5532[$-$4] && $-$0.08522 & 1.3379[$-$5] \\
BI    & $-$0.3987  & $-$3.6263[$-$4] &&  $-$0.09234   & 2.3667[$-$5]  && $-$0.3453 & 2.6411[$-$4] && $-$0.07971 & 1.2694[$-$5] \\
QED   & $-$0.3986  & $-$3.6161[$-$4] &&  $-$0.09254   & 2.3734[$-$5]  && $-$0.3453 & 2.6951[$-$4] && $-$0.07989 & 1.2716[$-$5] \\[0.2cm]
\multicolumn{12}{c}{\bf Zn$\bf ^{20+}$} \\
DHF   & $-$0.2611  & $-$1.3550[$-$6] &&  $-$0.3157    & 2.5105[$-$5]  && $-$0.2273 & 1.0671[$-$5] && $-$0.2717  & 1.1991[$-$6] \\
RCI   & $-$0.2604  & $-$1.3499[$-$6] &&  $-$0.3162    & 2.6076[$-$5]  && $-$0.2267 & 1.0933[$-$5] && $-$0.2721  & 1.2055[$-$6] \\
BI    & $-$0.2605  & $-$1.3601[$-$6] &&  $-$0.3161    & 2.6990[$-$5]  && $-$0.2269 & 1.0869[$-$5] && $-$0.2720  & 1.2136[$-$6] \\
QED   & $-$0.2605  & $-$1.3569[$-$6] &&  $-$0.3161    & 2.6995[$-$5]  && $-$0.2268 & 1.0865[$-$5] && $-$0.2721  & 1.2155[$-$6] \\
\hline
\hline
\end{tabular}
\end{table}

\begin{figure}
\centering
\includegraphics[scale=0.8]{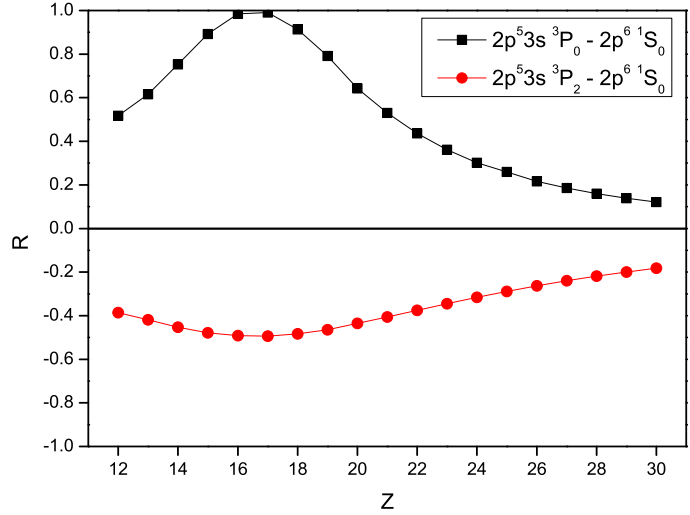}
\caption{\label{IE} The $R$ values for magnetic-field $2p^53s~^3P_{0,2} ~-~ 2p^6~^1S_0$ transition.}
\end{figure}

An earlier estimate of the MIT rate for Ne-like Ar have only included one perturbating state~\cite{Beiersdorfer2003}, but we find that this model is not sufficient and illustrate in Figure \ref{IE} the relative importance of the two included perturbers in Eq. (\ref{MIT-4}) through the ratio
\begin{equation}
\label{IR}
R^{\Gamma} = \frac{2 d^{\Gamma}_3 d^{\Gamma}_1 \langle 2p^6~^1S_0 || {\bf P}^{(1)} || 2p^53s~^3P_1 \rangle \langle 2p^6~^1S_0 || {\bf P}^{(1)} || 2p^53s~^1P_1 \rangle}{(d^{\Gamma}_3 \langle 2p^6~^1S_0 || {\bf P}^{(1)} || 2p^53s~^3P_1 \rangle)^2 + (d^{\Gamma}_1 \langle 2p^6~^1S_0 || {\bf P}^{(1)} || 2p^53s~^1P_1 \rangle)^2}
\end{equation}
as a function of the atomic number $Z$. The value of $R$ varies between 1 and $-$1, depending on the relative phase between two terms involved in the transition amplitudes of MITs. If $R > 0$, these two perturbers make constructive contribution to the MIT rate. Otherwise, there will be a cancelation between perturbers in the rate. When $|R| = 1$, two perturbers are of equivalent importance and thus must both be taken into account. As can be seen from Figure~\ref{IE}, the values of $R$ vary non-monotonously along the isoelectronic sequence for these two MITs, and are close to one, especially for the ions with $14 < Z < 20$. It is clear that the two contributions are of comparable size for all ions under investigation and thus both have to be included for accurate results. With respect to the sign of $R$, we find that the contribution from $2p^53s~^3P_1$ and $2p^53s~^1P_1$ perturbers are constructive for the magnetic field induced $2p^53s~^3P_0 ~-~ 2p^6~^1S_0$ E1 transition and are destructive for the magnetic field induced $2p^53s~^3P_2 ~-~ 2p^6~^1S_0$ E1 transition.


\subsection{Breit-Pauli estimates of MIT rates}
In this section, we confirm our Dirac-Hartree-Fock and Breit results by a Hatree-Fock (HF) calculation including relativistic corrections through the Breit-Pauli (BP) approximation for the magnetic field induced $2p^53s~^3P_0 ~-~ 2p^6~^1S_0$ transition rate of Mg$^{2+}$. Including the four main CSF's, the resulting atomic state wave functions labelled by the largest LSJ-component can be obtained with the ATSP2K package~\cite{FroeseFischer2007},
\begin{equation}
\begin{array}{lcl}
|``\,2p^5 3s~^3P_0" \rangle & = & | ^3P_0\rangle \\
|``\,2p^5 3s~^3P_1" \rangle & = & + 0.9738 | ^3P_1\rangle - 0.2275 | ^1P_1\rangle \\
|``\,2p^5 3s~^3P_2" \rangle & = & | ^3P_2\rangle \\
|``\,2p^5 3s~^1P_1" \rangle & = & + 0.9738 | ^1P_1\rangle + 0.2275 | ^3P_1\rangle \,.
\end{array}
\end{equation}
Using this basis, atomic parameters involved in the MIT such as the transition energies and line strengths, are calculated and listed in Table~\ref{BP}. The magnetic interaction Hamiltonian $H_m$ in the the non-relativistic approximation, assuming an magnetic field in the $z$-direction, is $H_m = \mu_B B(L_z+g_s S_z)$~\cite{Cowan1981}. The off-diagonal matrix elements between the CSFs can be expressed as~\cite{Jonsson2002}
\begin{equation}
\langle \gamma L S J M_J |L_z + g_s S_z| \gamma' L' S' J' M_J \rangle = \delta_{\gamma \gamma'} \delta_{L L'} \delta_{S S'} \delta_{J, J'(=J-1)} g_{J, J'(=J-1)}(LS) (J^2 - M^2)^{1/2},
\end{equation}
where
\begin{equation}
g_{J, J-1}(LS) = -(g_s - 1) \times \sqrt{\frac{(J+L+S+1)(J+L-S)(J+S-L)(L+S-J+1)}{4J^2(2J-1)(2J+1)}},
\end{equation}
with $g_s=2.00232$.
Hence we can write the matrix elements in the relation for the reduced mixing coefficients $d^R_{\mathcal{S}}$ (see Eq. (\ref{MC}) and (\ref{WF-1})) as
\begin{equation} \label{BPMC}
d^R_{\mathcal{S}} = \mu_B \frac{\langle ``2p^53s~^{\mathcal{S}}P_1" |L_z + g_sS_z | ``2p^53s~^3P_0" \rangle}{E(``2p^53s~^3P_0") - E(``2p^53s~^{\mathcal{S}}P_1")} = c(^3P_1) \frac{\mu_B}{\Delta E} g_{J,J-1}(LS)
\end{equation}
where $c(^3P_1)$ is the content of $^3P_1$ in the $``2p^53s~^3P_1"$ and $``2p^53s~^1P_1"$ states, and $\Delta E = E(``2p^53s~^3P_0") - E(``2p^53s~^{\mathcal{S}}P_1")$ and $L=S=J=1$ for the case under investigation. The resulting mixing coefficients can be found in Table~\ref{BP}.

As can be seen from Table~\ref{BP}, all atomic parameters obtained with Hartree-Fock calculations including relativistic corrections through Breit-Pauli approximation are in good agreement with Dirac-Hartree-Fock calculations with the Breit interaction. Substituting these atomic data into Eq. (\ref{MIT-4}), we obtain the reduced rate, $A^R_{MIT}$ = 82 $s^{-1}$T$^{-2}$, for the $2p^53s~^3P_0 ~-~ 2p^6~^1S_0$ MIT in the case of Ne-like Mg, which is in good consistence with the full relativistic calculation.

\begin{table}
\caption{\label{BP} The reduced magnetic field induced $2p^53s~^3P_0 ~-~ 2p^6~^1S_0$ E1 transition rate $A^R_{MIT}$ (in s$^{-1}$T$^{-2}$) in conjunction with the reduced mixing coefficients $d^R_{\mathcal{S}}$ (in T$^{-1}$), the reduced matrix element $\langle 2p^6~^1S_0 || {\bf P}^{(1)} || 2p^53s~^{\mathcal{S}}P_1 \rangle$ in the length gauge of the electric dipole transition ${\bf P}^{(1)}$ operator (in a.u.) and the wavelength $\lambda$ (in \AA) of the MIT for Ne-like Mg. The values marked with HF-BP and DHF-BI are obtained from a Hartree-Fock calculation including relativistic corrections through Breit-Pauli approximation and a Dirac-Hartree-Fock calculation with the Breit interaction, respectively. Numbers in square brackets are the power of 10.}
\begin{tabular}{ccccccccccccccccccccccccccc}
\hline
\hline
Method & $d^R_3$  & $\langle 2p^6~^1S_0 || {\bf P}^{(1)} || 2p^53s~^3P_1 \rangle$ && $d^R_1$  &  $\langle 2p^6~^1S_0 || {\bf P}^{(1)} || 2p^53s~^1P_1 \rangle$ & $\lambda$ &  $A^R_{MIT}$   \\
\hline
HF-BP  & $-$3.801[$-$4] &    $-$0.0865   &&   2.351[$-$5]  &  0.370  &  239  & 82 \\
DHF-BI & $-$3.626[$-$4] &    $-$0.0844   &&   2.367[$-$5]  &  0.371  &  239  & 78 \\
\hline
\hline
\end{tabular}
\end{table}

\subsection{The estimates of uncertainties}
There are two main error sources in the present calculations. One is related to the quality of atomic state wave functions, and the other is the perturbative treatment of the interaction between an external magnetic field and an atomic system. For the former the uncertainties result from the neglected electron correlations and other effects such as the frequency-dependent Breit interaction (FDBI). We estimate that the residual correlations and the FDBI effects contribute to the off-diagonal matrix elements and the mixing coefficients by about 3\%. The perturbative treatment of the magnetic interaction brings about fractional errors which are much less than 1\% for the MIT rates. Therefore, the total uncertainties in the MIT rates are less than 5\%.

\section{Results and discussions}
\subsection{The magnetic-field induced $2p^53s~^3P_0 ~-~ 2p^6~^1S_0$ E1 transition}

The reduced MIT rates (see Eq. (\ref{RMIT})) are reported in Table~\ref{MIT-1a}, with inclusion of only the $^1P_1$ perturber, as well as of both $^1P_1$ and $^3P_1$. It is clear that the contribution from the $2p^53s~^3P_1$ is indeed significant to the MIT probabilities for a major part of the sequence. Using the reduced total MIT rates one can readily obtain the MIT probabilities by Eq. (\ref{RMIT}) for any magnetic field, but we give the values for some examples in Table~\ref{MIT-1b}. With respect to plasma diagnostics, only the rate for the most abundant isotope of Ne-like ions without nuclear spin are given. In Ref.~\cite{Beiersdorfer2003} Beiersdorfer \textit{et al.} calculated the MIT rate for Ne-like Ar and obtained a value of 2440 s$^{-1}$ in the case of $B=3$ T. Using the same magnetic field strength, we have a MIT rate of 3004 s$^{-1}$. The difference between their result and ours are surprisingly small considering the fact that they neglected the $2p^53s~^3P_1$ perturber. It can be seen from Table~\ref{MIT-1a} that the inclusion of this perturber increases the MIT rate by a factor of 3 for Ne-like Ar. We believe that other uncertainties in their calculations cancel the effect of neglecting one perturber.

\begin{table}[!ht]
\caption{\label{MIT-1a} Reduced magnetic field induced $2p^53s~^3P_0 ~-~ 2p^6~^1S_0$ E1 transition probabilities $A_{MIT}^R$ (in s$^{-1}$T$^{-2}$) for Ne-like ions with $12 \le Z \le 30$. $A^{R}_{MIT}$ are obtained by inclusion of only $2p^53s~^1P_1$ perturber (labelled "only $^1P_1$") and of both the $2p^53s~^1P_1$ and the $2p^53s~^3P_1$ perturbers (labelled "Total"), respectively. For comparison, $2p^53s~^3P_0 ~-~ 2p^53s~^3P_1$ M1 transition probabilities (in $s^{-1}$) are displayed. The numbers in square brackets represent the power of 10. MIT rates for any magnetic field can be obtained with $A_{MIT}^R$ through $A_{MIT} = A_{MIT}^R B^2$.}
\begin{tabular}{lccclccccccccccc}
\hline
\hline
Ions &  only $^1P_1$  &  Total  & M1 && Ions & only $^1P_1$ & Total & M1\\
\hline
Mg$^{2+}$  & 5    & 105 & 5.58[-2] &&   Ti$^{12+}$ & 369  & 558  & 1.35[3] \\
Al$^{3+}$  & 9    & 135 & 2.14[-1] &&   V$^{13+}$  & 443  & 624  & 2.64[3] \\
Si$^{4+}$  & 16   & 168 & 7.64[-1] &&   Cr$^{14+}$ & 521  & 695  & 4.96[3] \\
P$^{5+}$   & 30   & 204 & 2.52     &&   Mn$^{15+}$ & 604  & 774  & 9.00[3] \\
S$^{6+}$   & 51   & 244 & 7.68     &&   Fe$^{16+}$ & 691  & 850  & 1.59[4] \\
Cl$^{7+}$  & 82   & 287 & 21.6     &&   Co$^{17+}$ & 783  & 936  & 2.73[4] \\
Ar$^{8+}$  & 123  & 334 & 56.1     &&   Ni$^{18+}$ & 879  & 1026 & 4.57[4] \\
K$^{9+}$   & 173  & 384 & 136      &&   Cu$^{19+}$ & 983  & 1125 & 7.47[4] \\
Ca$^{10+}$ & 235  & 438 & 308      &&   Zn$^{20+}$ & 1091 & 1228 & 1.20[5] \\
Sc$^{11+}$ & 300  & 496 & 661      &&                            &         \\
\hline
\hline
\end{tabular}
\end{table}

\begin{table}[!ht]
\caption{\label{MIT-1b} Magnetic field induced $2p^53s~^3P_0 ~-~ 2p^6~^1S_0$ E1 transition probabilities $A_{MIT}$ (in s$^{-1}$) for abundant isotopes without nuclear spin of Ne-like ions between Mg III and Zn XXI. $A_{MIT}$ are given for magnetic fields $B=0.5, 1.5$ and 2.5 T through Eq. (\ref{RMIT}). Magnetic dipole $^3P_0 ~-~ ^3P_1$ transitions rates $A_{M1}$ (in s$^{-1}$) are presented for convenience.}
\begin{tabular}{cccccccccccccccc}
\hline
\hline
     & \multicolumn{3}{c}{$A_{MIT}$}  \\
\cline{2-4}Ions &   $B=0.5$  &  $B=1.5$   &   $B=2.5$  & $A_{M1}$ \\
\hline
$^{24}$Mg$^{2+}$  &  26  & 235  & 653  & 0.0558 \\
$^{28}$Si$^{4+}$  &  42  & 378  & 1049 & 0.764  \\
$^{32}$S$^{6+}$   &  61  & 549  & 1526 & 7.68   \\
$^{40}$Ar$^{8+}$  &  83  & 751  & 2086 & 56.1   \\
$^{40}$Ca$^{10+}$ &  109 & 985  & 2736 & 308    \\
$^{48}$Ti$^{12+}$ &  140 & 1256 & 3488 & 1349   \\
$^{52}$Cr$^{14+}$ &  174 & 1563 & 4341 & 4955   \\
$^{56}$Fe$^{16+}$ &  212 & 1912 & 5312 & 15880  \\
$^{58}$Ni$^{18+}$ &  257 & 2309 & 6415 & 45672  \\
$^{64}$Zn$^{20+}$ &  307 & 2764 & 7677 & 120249 \\
\hline
\hline
\end{tabular}
\end{table}

\begin{figure}[!ht]
\centering
\includegraphics[scale=0.8]{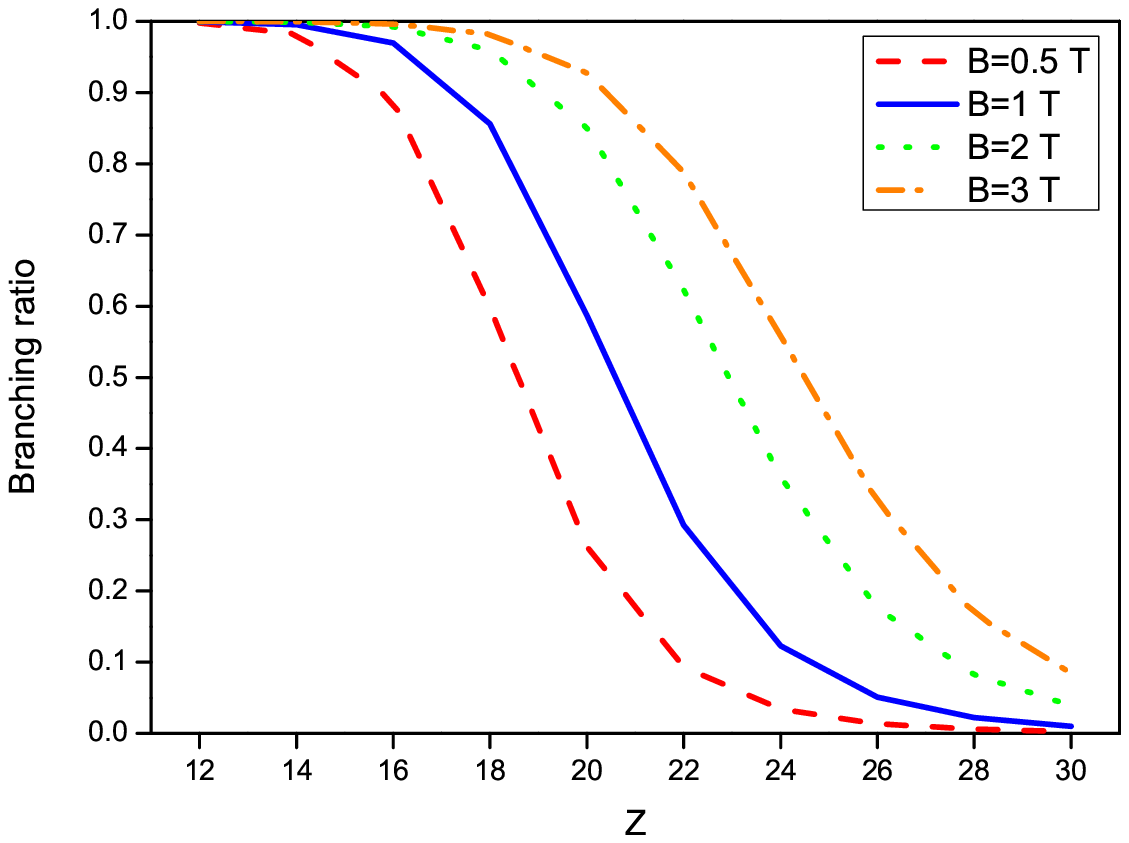}
\caption{\label{BR} The branching ratio of MIT for the $2p^53s~^3P_0$ state under circumstances of $B=0.5, 1, 2, 3$ T, respectively. }
\end{figure}

\begin{figure}[!ht]
\centering
\includegraphics[scale=0.8]{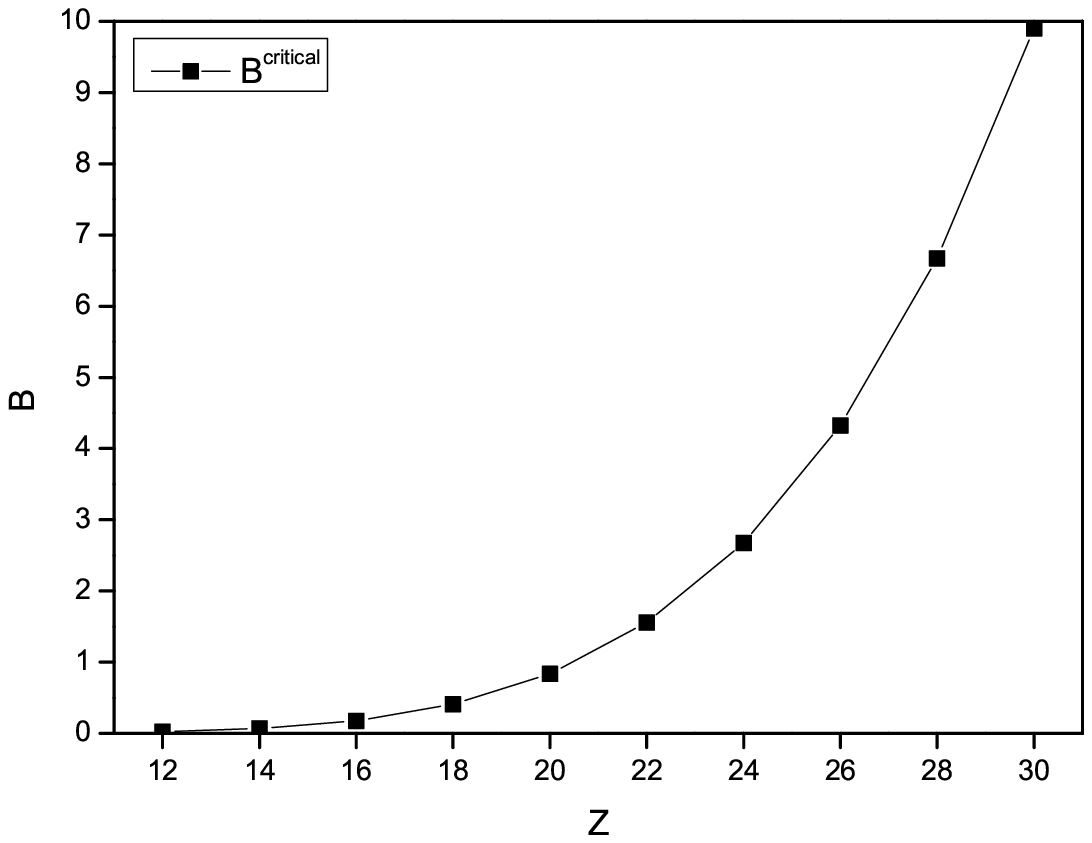}
\caption{\label{cB} The strength of critical magnetic fields B$^{critical}$ in T (see Eq. (\ref{Bcritical})).}
\end{figure}

In the last column of Table~\ref{MIT-1b} we also present the rates of the $2p^53s~^3P_0 ~-~ 2p^53s~^3P_1$ M1 transition. It is clear that the MIT is a dominant decay channel of the $2p^53s~^3P_0$ state for low-$Z$ ions even in relatively weak external magnetic fields. The branching ratios of the MIT channel ($\frac{A_{MIT}}{A_{MIT}+A_{M1}}$) along the Ne-like isoelectronic sequence are given in Figure~\ref{BR} for some magnetic fields. This property is also related to the relative intensity between the MIT line and the $2p^53s~^1P_1 ~-~ 2p^6~^1S_0$ E1 line (3F)~\cite{Beiersdorfer2003}. Regarding plasma diagnostics it is interesting to note for which magnetic field the MIT rate equals the M1 rate. The corresponding critical magnetic field strength can be calculated using
\begin{equation}
\label{Bcritical}
B^{critical} = \sqrt{\frac{A_{M1}}{A^{R}_{MIT}}}.
\end{equation}
This field is depicted in Figure~\ref{cB} as a function of $Z$. As can be seen from this figure, the magnetic field covers a wide range from a few thousand G to several T between Mg III and Zn XXI.

\subsection{The magnetic-field induced $2p^53s~^3P_2 ~-~ 2p^6~^1S_0$ E1 transition}
Resembling the magnetic-field induced $2p^53s~^3P_0 ~-~ 2p^6~^1S_0$ E1 transition, one-photon E1 decay channels are opened by the external magnetic field for the $2p^53s~^3P_2$ state with $M = +1, \, 0, \, \mbox{and} \, -1$. The magnetic-field induced $2p^53s~^3P_2 ~-~ 2p^6~^1S_0$ transition rates are dependent on the magnetic quantum number $M$ of sublevels belonging to the $2p^53s~^3P_2$ state through the mixing coefficients $d_{\mathcal{S}}$ in Eq. (\ref{MC}). For the $2p^53s~^3P_2$ state, $d_{\mathcal{S}}$ is given by
\begin{eqnarray}
d_{\mathcal{S}}(M) &=& \frac{\langle ~^{\mathcal{S}}P_1 M | H_m | ^3P_2 M \rangle}{E(^3P_2) - E(^{\mathcal{S}}P_1)} \nonumber \\
                     &=& B \sqrt{\frac{4-M^2}{6}} \frac{\langle ^3P_2 || {\bf N}^{(1)} + \Delta {\bf N}^{(1)} || ~^{\mathcal{S}}P_1 \rangle}{E(^3P_2) - E(^{\mathcal{S}}P_1)}.
\end{eqnarray}
As a result, the rates of the $2p^53s~^3P_2 ~-~ 2p^6~^1S_0$ MITs for individual sublevels can be expressed as
\begin{eqnarray}
A_{MIT}(M) &=& \frac{2.02613 \times 10^{18}}{3 \lambda^3} \frac{B^2(4-M^2)}{6} \left |\sum_{\mathcal{S}(=3,1)}\frac{\langle ^3P_2 || {\bf N}^{(1)} + \Delta {\bf N}^{(1)} || ~^{\mathcal{S}}P_1 \rangle}{E(^3P_2) - E(^{\mathcal{S}}P_1)} \langle ^1S_0 || {\bf P}^{(1)} || ~^{\mathcal{S}}P_1 \rangle \right|^2. \nonumber \\
\end{eqnarray}
The two equations above show that the definition of the reduced MIT rate (see Eq. (\ref{RMIT})) is still valid for this transition. Furthermore, $A_{MIT}(M)$ is proportional to $(4-M^2) B^2$. In Table~\ref{MIT-2} we report the reduced MIT rates from each magnetic sublevels in the $2p^53s~^3P_2$ state to the ground state, which can be used to calculate the MIT rates in a certain magnetic field.

\begin{table}[!ht]
\caption{\label{MIT-2} Reduced magnetic-field induced $2p^53s~^3P_2 ~-~ 2p^6~^1S_0$ E1 transition probabilities $A_{MIT}^{R}(M)$ (in s$^{-1}$T$^{-2}$) for Ne-like ions with $12 \le Z \le 30$. $A^{R}_{MIT}(M)$ are obtained by inclusion of both the $2p^53s~^1P_1$ and the $2p^53s~^3P_1$ perturbers. $M$ represents the magnetic quantum numbers of sublevels in the $2p^53s~^3P_2$ state. MIT rates for any magnetic field can be obtained with $A_{MIT}^R(M)$ through $A_{MIT}(M) = A_{MIT}^R(M) B^2$}
\begin{tabular}{lccclccccccccccc}
\hline
\hline
Ions       & $M = 0$   & $M = \pm 1$   &&  Ions        &  $M = 0$   & $M = \pm 1$ \\
\hline
Mg$^{2+}$  &    15     &    11         &&  Ti$^{12+}$  &   67       &   50          \\
Al$^{3+}$  &    18     &    14         &&  V$^{13+}$   &   74       &   55          \\
Si$^{4+}$  &    22     &    16         &&  Cr$^{14+}$  &   81       &   61          \\
P$^{5+}$   &    26     &    19         &&  Mn$^{15+}$  &   89       &   67          \\
S$^{6+}$   &    31     &    23         &&  Fe$^{16+}$  &   98       &   73          \\
Cl$^{7+}$  &    36     &    27         &&  Co$^{17+}$  &   106      &   80          \\
Ar$^{8+}$  &    41     &    31         &&  Ni$^{18+}$  &   114      &   86          \\
K$^{9+}$   &    47     &    35         &&  Cu$^{19+}$  &   124      &   93          \\
Ca$^{10+}$ &    53     &    40         &&  Zn$^{20+}$  &   134      &   100         \\
Sc$^{11+}$ &    60     &    45         &&   \\
\hline
\hline
\end{tabular}
\end{table}

In absence of an external magnetic field, the magnetic quadrupole (M2) $2p^53s~^3P_2 ~-~ 2p^6~^1S_0$ transition is mainly a one-photon decay channel. Hence the lifetime for the $2p^53s~^3P_2$ state is given by
\begin{equation}
\label{t1}
\tau = \frac{1}{A_{M2}}.
\end{equation}
When the external magnetic field is introduced, the MIT should be taken into account and thus the lifetimes is dependent on the $M$ of sublevels in $2p^53s~^3P_2$ state. For each sublevels, we have
\begin{equation}
\label{t2}
\tau(M) = \frac{1}{A_{M2} + A_{MIT}(M)}.
\end{equation}
Using Eq. (\ref{t1}) and Eq. (\ref{t2}), we compute the MIT rates of individual sublevels in the $2p^53s~^3P_2$ state under circumstances of $B=0.5$, 1.5, and 2.5 T, respectively, for most abundant isotopes without nuclear spin. The results are presented in Table~\ref{LT} as well as the $2p^53s~^3P_2 ~-~ 2p^6~^1S_0$ M2 transition rates with $B=0$ T. In addition, we also display statistical average values $\overline{\tau}$ of lifetimes of the $2p^53s~^3P_2$ state in the external magnetic field, which are obtained by
\begin{equation}
\label{alt}
\overline{\tau} = \frac{2J+1}{\sum (A_{M2} + A_{MIT}(M))},
\end{equation}
where the summation is made over MIT and M2 decay channels from all magnetic sublevels in the $2p^53s~^3P_2$ state. As can be seen from Table~\ref{LT}, the external magnetic field little influences the lifetime of the $2p^53s~^3P_2$ due to the destructive contributions from $^3P_1$ and $^1P_1$ perturbers and relatively large M2 transition rates, but still generates observable effect for ions near the neutral end. For instance, the external magnetic field reduces the lifetime by 90\% in a $B=2.5$ T magnetic field for Mg III.

\begin{table}
\scriptsize{
\caption{\label{LT} $2p^53s~^3P_2 ~-~ 2p^6~^1S_0$ M2 transition rates $A_{M2}$ (in s$^{-1}$) and magnetic field induced $2p^53s~^3P_2 ~-~ 2p^6~^1S_0$ E1 transition rates $A_{MIT}(M)$ (in s$^{-1}$) for each of the magnetic sublevels $M$ in the $2p^53s~^3P_2$ state under circumstances of $B$=0.5, 1.5 and 2.5 T in Ne-like ions. The average lifetime $\overline{\tau}$ (in s) for the $2p^53s~^3P_2$ state in the magnetic fields are given. Numbers in square brackets are the power of 10.}
\begin{tabular}{ccccccccccccccccccccccc}
\hline
\hline
  &  \multicolumn{1}{c}{$B=0$ T} && \multicolumn{3}{c}{$B=0.5$ T} && \multicolumn{3}{c}{$B=1.5$ T} && \multicolumn{3}{c}{$B=2.5$ T} \\
\cline{2-2}\cline{4-6}\cline{8-10}\cline{12-14} ions & $A_{M2}$ && $A_{MIT}(0)$ & $A_{MIT}(\pm 1)$ & $\overline{\tau}$ && $A_{MIT}(0)$ & $A_{MIT}(\pm 1)$ & $\overline{\tau}$ && $A_{MIT}(0)$ & $A_{MIT}(\pm 1)$ & $\overline{\tau}$ \\
\hline
$^{24}$Mg$^{2+}$  & 7.63    && 3.722    & 2.791     & 1.054[-1] && 3.350[1] & 2.512[1]  & 4.102[-2] && 9.304[1] & 6.978[1]  & 1.847[-2] \\
$^{28}$Si$^{4+}$  & 1.12[2] && 5.434    & 4.076     & 8.695[-3] && 4.891[1] & 3.668[1]  & 7.313[-3] && 1.359[2] & 1.019[2]  & 5.549[-3] \\
$^{32}$S$^{6+}$   & 7.64[2] && 7.645    & 5.733     & 1.303[-3] && 6.880[1] & 5.160[1]  & 1.253[-3] && 1.911[2] & 1.433[2]  & 1.164[-3] \\
$^{40}$Ar$^{8+}$  & 3.45[3] && 1.027[1] & 7.700     & 2.892[-4] && 9.240[1] & 6.930[1]  & 2.858[-4] && 2.567[2] & 1.925[2]  & 2.792[-4] \\
$^{40}$Ca$^{10+}$ & 1.20[4] && 1.324[1] & 9.931     & 8.306[-5] && 1.192[2] & 8.938[1]  & 8.269[-5] && 3.310[2] & 2.483[2]  & 8.198[-5] \\
$^{48}$Ti$^{12+}$ & 3.52[4] && 1.663[1] & 1.247[1]  & 2.838[-5] && 1.497[2] & 1.122[2]  & 2.833[-5] && 4.157[2] & 3.118[2]  & 2.822[-5] \\
$^{52}$Cr$^{14+}$ & 9.01[4] && 2.035[1] & 1.527[1]  & 1.110[-5] && 1.832[2] & 1.374[2]  & 1.109[-5] && 5.088[2] & 3.816[2]  & 1.107[-5] \\
$^{56}$Fe$^{16+}$ & 2.08[5] && 2.440[1] & 1.830[1]  & 4.806[-6] && 2.196[2] & 1.647[2]  & 4.803[-6] && 6.100[2] & 4.575[2]  & 4.799[-6] \\
$^{58}$Ni$^{18+}$ & 4.41[5] && 2.861[1] & 2.146[1]  & 2.268[-6] && 2.575[2] & 1.931[2]  & 2.268[-6] && 7.153[2] & 5.365[2]  & 2.266[-6] \\
$^{64}$Zn$^{20+}$ & 8.82[5] && 3.341[1] & 2.506[1]  & 1.134[-6] && 3.007[2] & 2.255[2]  & 1.133[-6] && 8.352[2] & 6.264[2]  & 1.133[-6] \\
\hline
\hline
\end{tabular}
}
\end{table}

\subsection{MITs in the case of Ne I}
The accurate determination of lifetimes for metastable states in the first excited configuration of rare gases is always appealing~\cite{Beck2002}. For neutral neon, there still exists an unresolved discrepancy in the lifetime of the $2p^53s~^3P_2$ state between experiment and theory. The theoretical value~\cite{Li2012} $\tau=17.63$ s differs from the measurement~\cite{Zinner2003} $14.73$ s by 20\%, which is much larger than the uncertainties in both theory and experiment. Therefore, we investigate the effect of an external magnetic field on the lifetime of this level. Regarding the strong electron correlations in neutral neon, we adopt a more complicated computational model for taking into account the correlation effects~\cite{Li2012}. In Table~\ref{Ne-1} the lifetime of the $2p^53s~^3P_2$ state for $^{20}$Ne is presented in cases of $B=0.01$, and 1 T, respectively. Comparing with the $2p^53s~^3P_2 ~-~ 2p^6~^1S_0$ M2 transition rate $A_{M2}$ in the case of $B=0$ T, we find from this table that the external magnetic field, even fairly weak, drastically reduces the lifetime of the $2p^53s~^3P_2$ state. In other words, the lifetime of the $2p^53s~^3P_2$ state is highly sensitive to the strength of the magnetic field. Hence the magnetic field effect on the $2p^53s~^3P_2$ level lifetime could be part of the reason behind the discrepancy.

\begin{table}
\caption{\label{Ne-1} $2p^53s~^3P_2 ~-~ 2p^6~^1S_0$ M2 transition rates $A_{M2}$ (in s$^{-1}$) and magnetic field induced $2p^53s~^3P_2 ~-~ 2p^6~^1S_0$ E1 transition rates $A_{MIT}(M)$ (in s$^{-1}$) for each of the magnetic sublevels ($M$) in the $2p^53s~^3P_2$ state for $^{20}$Ne without and with an external magnetic field of $B=0.01$ T and 1 T, respectively. $\overline{\tau}$ is an average level lifetime (in s) obtained with Eq. (\ref{alt}). Numbers in square brackets represent the power of 10.}
\begin{tabular}{ccccccccccccccccccccccccc}
\hline
\hline
  &  \multicolumn{1}{c}{$B=0$ T} && \multicolumn{3}{c}{$B=0.01$ T} && \multicolumn{3}{c}{$B=1$ T}  \\
\cline{2-2}\cline{4-6}\cline{8-10} ions & $A_{M2}$ && $A_{MIT}(0)$ & $A_{MIT}(\pm 1)$  & $\overline{\tau}$ && $A_{MIT}(0)$ & $A_{MIT}(\pm 1)$ & $\overline{\tau}$ \\
\hline
$^{20}$Ne & 5.672[$-$2] && 1.197[$-$3] & 8.979[$-$4]  & 17.45 && 11.97 & 8.979 & 1.655[$-$1] \\
\hline
\hline
\end{tabular}
\end{table}

In addition, we predict the magnetic field induced $2p^53s~^3P_0 ~-~ 2p^6~^1S_0$ E1 transition rate for neutral neon. This value reaches 95.2 s$^{-1}$ in an 1 T external magnetic field, which is much larger than the M1 rate, $A_{M1} = 2.358 \times 10^{-3}$s$^{-1}$. Moreover, the MIT rate is still comparable to the M1 even for a field of only 0.005 T.

\section{Summary}
To conclude, we have predicted rates for the magnetic field induced (MIT) $2p^53s~^3P_{2,0} ~-~ 2p^6~^1S_0$ E1 transitions using the MCDHF method for Ne-like ions between Mg III and Zn XXI without nuclear spin. We emphasize that both $2p^53s~^1P_1$ and $2p^53s~^3P_1$ perturber states are essential to include in order to obtain reliable the MIT rates. Using the reduced MIT rates, $A_{MIT}^R$, reported in this paper, it is possible to predict rates for any magnetic field strength. The atomic data presented in this work can be utilized for modelling plasma spectra. One should keep in mind, however, there often exists the angular distribution of intensity for emission lines from plasma especially in the presence of an external magnetic field. The relevant works are ongoing.

We investigate the competition of the MITs with other possible one-photon decay channels. It is found that the magnetic field induced $2p^53s~^3P_0 ~-~ 2p^6~^1S_0$ E1 transition is the dominant decay channel for low-$Z$ ions compared to the $2p^53s~^3P_0 ~-~ 2p^53s~^3P_1$ M1 transition, while the influence of magnetic fields on the lifetime of the $2p^53s~^3P_2$ state is small but still observable for the ions at the neutral end of the sequence.

In order to help resolve the discrepancy in the lifetime of the $2p^53s~^3P_2$ state for neutral neon between experiment and theory, the MIT rates in $^{20}$Ne are calculated as well. We find that the lifetime of both $2p^53s~^3P_2$ and $2p^53s~^3P_0$ states are extremely sensitive to the strength of magnetic fields. Dependent on the experiment setup the effect of any magnetic field could thus be a possible reason for this unresolved discrepancy.


\begin{acknowledgments}
This work was supported by the National Natural Science Foundation of China No. 11074049, and by Shanghai Leading Academic Discipline Project No. B107. We gratefully acknowledge support from the Swedish Research Council (Vetenskapsr{\aa}det) and the Swedish Institute under the Visby-program.
\end{acknowledgments}


\bibliography{MIT,Ne-like}

\begin{thebibliography}{22}%
\makeatletter
\providecommand \@ifxundefined [1]{%
 \@ifx{#1\undefined}
}%
\providecommand \@ifnum [1]{%
 \ifnum #1\expandafter \@firstoftwo
 \else \expandafter \@secondoftwo
 \fi
}%
\providecommand \@ifx [1]{%
 \ifx #1\expandafter \@firstoftwo
 \else \expandafter \@secondoftwo
 \fi
}%
\providecommand \natexlab [1]{#1}%
\providecommand \enquote  [1]{``#1''}%
\providecommand \bibnamefont  [1]{#1}%
\providecommand \bibfnamefont [1]{#1}%
\providecommand \citenamefont [1]{#1}%
\providecommand \href@noop [0]{\@secondoftwo}%
\providecommand \href [0]{\begingroup \@sanitize@url \@href}%
\providecommand \@href[1]{\@@startlink{#1}\@@href}%
\providecommand \@@href[1]{\endgroup#1\@@endlink}%
\providecommand \@sanitize@url [0]{\catcode `\\12\catcode `\$12\catcode
  `\&12\catcode `\#12\catcode `\^12\catcode `\_12\catcode `\%12\relax}%
\providecommand \@@startlink[1]{}%
\providecommand \@@endlink[0]{}%
\providecommand \url  [0]{\begingroup\@sanitize@url \@url }%
\providecommand \@url [1]{\endgroup\@href {#1}{\urlprefix }}%
\providecommand \urlprefix  [0]{URL }%
\providecommand \Eprint [0]{\href }%
\providecommand \doibase [0]{http://dx.doi.org/}%
\providecommand \selectlanguage [0]{\@gobble}%
\providecommand \bibinfo  [0]{\@secondoftwo}%
\providecommand \bibfield  [0]{\@secondoftwo}%
\providecommand \translation [1]{[#1]}%
\providecommand \BibitemOpen [0]{}%
\providecommand \bibitemStop [0]{}%
\providecommand \bibitemNoStop [0]{.\EOS\space}%
\providecommand \EOS [0]{\spacefactor3000\relax}%
\providecommand \BibitemShut  [1]{\csname bibitem#1\endcsname}%
\let\auto@bib@innerbib\@empty
\bibitem [{\citenamefont {Johns-Krull}\ \emph {et~al.}(1999)\citenamefont
  {Johns-Krull}, \citenamefont {Valenti},\ and\ \citenamefont
  {Koresko}}]{Johns-Krull1999}%
  \BibitemOpen
  \bibfield  {author} {\bibinfo {author} {\bibfnamefont {C.~M.}\ \bibnamefont
  {Johns-Krull}}, \bibinfo {author} {\bibfnamefont {J.~A.}\ \bibnamefont
  {Valenti}}, \ and\ \bibinfo {author} {\bibfnamefont {C.}~\bibnamefont
  {Koresko}},\ }\href@noop {} {\bibfield  {journal} {\bibinfo  {journal} {ApJ}\
  }\textbf {\bibinfo {volume} {516}},\ \bibinfo {pages} {900} (\bibinfo {year}
  {1999})}\BibitemShut {NoStop}%
\bibitem [{\citenamefont {Adams}\ \emph {et~al.}(2002)\citenamefont {Adams},
  \citenamefont {Lee}, \citenamefont {Scott}, \citenamefont {Chung},\ and\
  \citenamefont {Klein}}]{Adams2002}%
  \BibitemOpen
  \bibfield  {author} {\bibinfo {author} {\bibfnamefont {M.~L.}\ \bibnamefont
  {Adams}}, \bibinfo {author} {\bibfnamefont {R.~W.}\ \bibnamefont {Lee}},
  \bibinfo {author} {\bibfnamefont {H.~A.}\ \bibnamefont {Scott}}, \bibinfo
  {author} {\bibfnamefont {H.~K.}\ \bibnamefont {Chung}}, \ and\ \bibinfo
  {author} {\bibfnamefont {L.}~\bibnamefont {Klein}},\ }\href {\doibase
  10.1103/PhysRevE.66.066413} {\bibfield  {journal} {\bibinfo  {journal}
  {Physical Review E}\ }\textbf {\bibinfo {volume} {66}},\ \bibinfo {pages}
  {066413} (\bibinfo {year} {2002})}\BibitemShut {NoStop}%
\bibitem [{\citenamefont {Stambulchik}\ \emph {et~al.}(2007)\citenamefont
  {Stambulchik}, \citenamefont {Tsigutkin},\ and\ \citenamefont
  {Maron}}]{Stambulchik2007}%
  \BibitemOpen
  \bibfield  {author} {\bibinfo {author} {\bibfnamefont {E.}~\bibnamefont
  {Stambulchik}}, \bibinfo {author} {\bibfnamefont {K.}~\bibnamefont
  {Tsigutkin}}, \ and\ \bibinfo {author} {\bibfnamefont {Y.}~\bibnamefont
  {Maron}},\ }\href {\doibase 10.1103/PhysRevLett.98.225001} {\bibfield
  {journal} {\bibinfo  {journal} {Physical Review Letters}\ }\textbf {\bibinfo
  {volume} {98}},\ \bibinfo {pages} {225001} (\bibinfo {year}
  {2007})}\BibitemShut {NoStop}%
\bibitem [{\citenamefont {Andrew}\ \emph {et~al.}(1967)\citenamefont {Andrew},
  \citenamefont {Cowan},\ and\ \citenamefont {Giacchetti}}]{Andrew1967}%
  \BibitemOpen
  \bibfield  {author} {\bibinfo {author} {\bibfnamefont {K.~L.}\ \bibnamefont
  {Andrew}}, \bibinfo {author} {\bibfnamefont {R.~D.}\ \bibnamefont {Cowan}}, \
  and\ \bibinfo {author} {\bibfnamefont {A.}~\bibnamefont {Giacchetti}},\
  }\href@noop {} {\bibfield  {journal} {\bibinfo  {journal} {J. Opt. Soc. Am.}\
  }\textbf {\bibinfo {volume} {57}},\ \bibinfo {pages} {715} (\bibinfo {year}
  {1967})}\BibitemShut {NoStop}%
\bibitem [{\citenamefont {Wood}\ \emph {et~al.}(1968)\citenamefont {Wood},
  \citenamefont {Andrew},\ and\ \citenamefont {Cowan}}]{Wood1968}%
  \BibitemOpen
  \bibfield  {author} {\bibinfo {author} {\bibfnamefont {D.~R.}\ \bibnamefont
  {Wood}}, \bibinfo {author} {\bibfnamefont {K.~L.}\ \bibnamefont {Andrew}}, \
  and\ \bibinfo {author} {\bibfnamefont {R.~D.}\ \bibnamefont {Cowan}},\
  }\href@noop {} {\bibfield  {journal} {\bibinfo  {journal} {J. Opt. Soc. Am.}\
  }\textbf {\bibinfo {volume} {58}},\ \bibinfo {pages} {830} (\bibinfo {year}
  {1968})}\BibitemShut {NoStop}%
\bibitem [{\citenamefont {Feldman}\ \emph {et~al.}(1967)\citenamefont
  {Feldman}, \citenamefont {Levitt}, \citenamefont {Manson},\ and\
  \citenamefont {Novick}}]{Feldman1967}%
  \BibitemOpen
  \bibfield  {author} {\bibinfo {author} {\bibfnamefont {P.}~\bibnamefont
  {Feldman}}, \bibinfo {author} {\bibfnamefont {M.}~\bibnamefont {Levitt}},
  \bibinfo {author} {\bibfnamefont {S.}~\bibnamefont {Manson}}, \ and\ \bibinfo
  {author} {\bibfnamefont {R.}~\bibnamefont {Novick}},\ }\href@noop {}
  {\bibfield  {journal} {\bibinfo  {journal} {Physica}\ }\textbf {\bibinfo
  {volume} {33}},\ \bibinfo {pages} {278} (\bibinfo {year} {1967})}\BibitemShut
  {NoStop}%
\bibitem [{\citenamefont {Levitt}\ \emph {et~al.}(1971)\citenamefont {Levitt},
  \citenamefont {Novick},\ and\ \citenamefont {Feldman}}]{Levitt1971}%
  \BibitemOpen
  \bibfield  {author} {\bibinfo {author} {\bibfnamefont {M.}~\bibnamefont
  {Levitt}}, \bibinfo {author} {\bibfnamefont {R.}~\bibnamefont {Novick}}, \
  and\ \bibinfo {author} {\bibfnamefont {P.~D.}\ \bibnamefont {Feldman}},\
  }\href@noop {} {\bibfield  {journal} {\bibinfo  {journal} {Phys. Rev. A}\
  }\textbf {\bibinfo {volume} {3}},\ \bibinfo {pages} {130} (\bibinfo {year}
  {1971})}\BibitemShut {NoStop}%
\bibitem [{\citenamefont {Beiersdorfer}\ \emph {et~al.}(2003)\citenamefont
  {Beiersdorfer}, \citenamefont {Scofield},\ and\ \citenamefont
  {Osterheld}}]{Beiersdorfer2003}%
  \BibitemOpen
  \bibfield  {author} {\bibinfo {author} {\bibfnamefont {P.}~\bibnamefont
  {Beiersdorfer}}, \bibinfo {author} {\bibfnamefont {J.~H.}\ \bibnamefont
  {Scofield}}, \ and\ \bibinfo {author} {\bibfnamefont {A.~L.}\ \bibnamefont
  {Osterheld}},\ }\href {\doibase 10.1103/PhysRevLett.90.235003} {\bibfield
  {journal} {\bibinfo  {journal} {Phys. Rev. Lett.}\ }\textbf {\bibinfo
  {volume} {90}},\ \bibinfo {pages} {235003} (\bibinfo {year}
  {2003})}\BibitemShut {NoStop}%
\bibitem [{\citenamefont {Li}\ \emph {et~al.}(2012)\citenamefont {Li},
  \citenamefont {J\"{o}nsson}, \citenamefont {Godefroid}, \citenamefont
  {Dong},\ and\ \citenamefont {Gaigalas}}]{Li2012}%
  \BibitemOpen
  \bibfield  {author} {\bibinfo {author} {\bibfnamefont {J.~G.}\ \bibnamefont
  {Li}}, \bibinfo {author} {\bibfnamefont {P.}~\bibnamefont {J\"{o}nsson}},
  \bibinfo {author} {\bibfnamefont {M.}~\bibnamefont {Godefroid}}, \bibinfo
  {author} {\bibfnamefont {C.~Z.}\ \bibnamefont {Dong}}, \ and\ \bibinfo
  {author} {\bibfnamefont {G.}~\bibnamefont {Gaigalas}},\ }\href {\doibase
  10.1103/PhysRevA.86.052523} {\bibfield  {journal} {\bibinfo  {journal}
  {Physical Review A}\ }\textbf {\bibinfo {volume} {86}},\ \bibinfo {pages}
  {052523} (\bibinfo {year} {2012})}\BibitemShut {NoStop}%
\bibitem [{\citenamefont {Andersson}\ \emph {et~al.}()\citenamefont
  {Andersson}, \citenamefont {Brage},\ and\ \citenamefont {Zou}}]{Andersson}%
  \BibitemOpen
  \bibfield  {author} {\bibinfo {author} {\bibfnamefont {M.}~\bibnamefont
  {Andersson}}, \bibinfo {author} {\bibfnamefont {T.}~\bibnamefont {Brage}}, \
  and\ \bibinfo {author} {\bibfnamefont {Y.}~\bibnamefont {Zou}},\ }\href@noop
  {} {\bibinfo  {journal} {Phys. Rev. A}\ ,\ \bibinfo {pages} {to be
  submitted}}\BibitemShut {NoStop}%
\bibitem [{\citenamefont {J\"{o}nsson}\ \emph
  {et~al.}(2013{\natexlab{a}})\citenamefont {J\"{o}nsson}, \citenamefont
  {Gaigalas}, \citenamefont {Bieroń}, \citenamefont {{Froese Fischer}},\ and\
  \citenamefont {Grant}}]{Jonsson2013}%
  \BibitemOpen
\bibfield  {journal} {  }\bibfield  {author} {\bibinfo {author} {\bibfnamefont
  {P.}~\bibnamefont {J\"{o}nsson}}, \bibinfo {author} {\bibfnamefont
  {G.}~\bibnamefont {Gaigalas}}, \bibinfo {author} {\bibfnamefont
  {J.}~\bibnamefont {Bieroń}}, \bibinfo {author} {\bibfnamefont
  {C.}~\bibnamefont {{Froese Fischer}}}, \ and\ \bibinfo {author}
  {\bibfnamefont {I.~P.}\ \bibnamefont {Grant}},\ }\href@noop {} {\bibfield
  {journal} {\bibinfo  {journal} {Computer Physics Communications}\ ,\ \bibinfo
  {pages} {in presee}} (\bibinfo {year} {2013}{\natexlab{a}})}\BibitemShut
  {NoStop}%
\bibitem [{\citenamefont {Andersson}\ and\ \citenamefont
  {J\"{o}nsson}(2008)}]{Andersson2008}%
  \BibitemOpen
  \bibfield  {author} {\bibinfo {author} {\bibfnamefont {M.}~\bibnamefont
  {Andersson}}\ and\ \bibinfo {author} {\bibfnamefont {P.}~\bibnamefont
  {J\"{o}nsson}},\ }\href {\doibase 10.1016/j.cpc.2007.07.014} {\bibfield
  {journal} {\bibinfo  {journal} {Computer Physics Communications}\ }\textbf
  {\bibinfo {volume} {178}},\ \bibinfo {pages} {156} (\bibinfo {year}
  {2008})}\BibitemShut {NoStop}%
\bibitem [{\citenamefont {Grant}(2007)}]{Grant2007}%
  \BibitemOpen
  \bibfield  {author} {\bibinfo {author} {\bibfnamefont {I.~P.}\ \bibnamefont
  {Grant}},\ }\href@noop {} {\emph {\bibinfo {title} {{Relativistic Quantum
  Theory of Atoms and Molecules}}}}\ (\bibinfo  {publisher} {New York:
  Springer},\ \bibinfo {year} {2007})\BibitemShut {NoStop}%
\bibitem [{\citenamefont {Cheng}\ and\ \citenamefont
  {Childs}(1985)}]{Cheng1985}%
  \BibitemOpen
  \bibfield  {author} {\bibinfo {author} {\bibfnamefont {K.~T.}\ \bibnamefont
  {Cheng}}\ and\ \bibinfo {author} {\bibfnamefont {W.~J.}\ \bibnamefont
  {Childs}},\ }\href@noop {} {\bibfield  {journal} {\bibinfo  {journal} {Phys.
  Rev. A}\ }\textbf {\bibinfo {volume} {31}},\ \bibinfo {pages} {2775}
  (\bibinfo {year} {1985})}\BibitemShut {NoStop}%
\bibitem [{\citenamefont {Johnson}\ \emph {et~al.}(1995)\citenamefont
  {Johnson}, \citenamefont {Plante},\ and\ \citenamefont
  {Sapirstein}}]{Johnson1995}%
  \BibitemOpen
  \bibfield  {author} {\bibinfo {author} {\bibfnamefont {W.~R.}\ \bibnamefont
  {Johnson}}, \bibinfo {author} {\bibfnamefont {D.~R.}\ \bibnamefont {Plante}},
  \ and\ \bibinfo {author} {\bibfnamefont {J.}~\bibnamefont {Sapirstein}},\
  }\href@noop {} {\bibfield  {journal} {\bibinfo  {journal} {Advances in
  Atomic, Molecular and Optical Physics}\ }\textbf {\bibinfo {volume} {35}},\
  \bibinfo {pages} {225} (\bibinfo {year} {1995})}\BibitemShut {NoStop}%
\bibitem [{\citenamefont {J\"{o}nsson}\ \emph {et~al.}(2011)\citenamefont
  {J\"{o}nsson}, \citenamefont {Bengtsson}, \citenamefont {Ekman},
  \citenamefont {Gustafsson}, \citenamefont {Karlsson}, \citenamefont
  {Gaigalas}, \citenamefont {{Froese Fischer}}, \citenamefont {Kato},
  \citenamefont {Murakami}, \citenamefont {Sakaue}, \citenamefont {Hara},
  \citenamefont {Watanabe}, \citenamefont {Nakamura},\ and\ \citenamefont
  {Yamamoto}}]{Jonsson2011}%
  \BibitemOpen
  \bibfield  {author} {\bibinfo {author} {\bibfnamefont {P.}~\bibnamefont
  {J\"{o}nsson}}, \bibinfo {author} {\bibfnamefont {P.}~\bibnamefont
  {Bengtsson}}, \bibinfo {author} {\bibfnamefont {J.}~\bibnamefont {Ekman}},
  \bibinfo {author} {\bibfnamefont {S.}~\bibnamefont {Gustafsson}}, \bibinfo
  {author} {\bibfnamefont {L.~B.}\ \bibnamefont {Karlsson}}, \bibinfo {author}
  {\bibfnamefont {G.}~\bibnamefont {Gaigalas}}, \bibinfo {author}
  {\bibfnamefont {C.}~\bibnamefont {{Froese Fischer}}}, \bibinfo {author}
  {\bibfnamefont {D.}~\bibnamefont {Kato}}, \bibinfo {author} {\bibfnamefont
  {I.}~\bibnamefont {Murakami}}, \bibinfo {author} {\bibfnamefont {H.~A.}\
  \bibnamefont {Sakaue}}, \bibinfo {author} {\bibfnamefont {H.}~\bibnamefont
  {Hara}}, \bibinfo {author} {\bibfnamefont {T.}~\bibnamefont {Watanabe}},
  \bibinfo {author} {\bibfnamefont {N.}~\bibnamefont {Nakamura}}, \ and\
  \bibinfo {author} {\bibfnamefont {N.}~\bibnamefont {Yamamoto}},\ }\href@noop
  {} {\bibfield  {journal} {\bibinfo  {journal} {National Institute for Fusion
  Science, Research Report}\ }\textbf {\bibinfo {volume} {NIFS-DATA-}},\
  \bibinfo {pages} {1} (\bibinfo {year} {2011})}\BibitemShut {NoStop}%
\bibitem [{\citenamefont {J\"{o}nsson}\ \emph
  {et~al.}(2013{\natexlab{b}})\citenamefont {J\"{o}nsson}, \citenamefont
  {Bengtsson}, \citenamefont {Ekman}, \citenamefont {Gustafsson}, \citenamefont
  {Karlsson}, \citenamefont {Gaigalas}, \citenamefont {{Froese Fischer}},
  \citenamefont {Kato}, \citenamefont {Murakami}, \citenamefont {Sakaue},
  \citenamefont {Hara}, \citenamefont {Watanabe}, \citenamefont {Nakamura},\
  and\ \citenamefont {Yamamoto}}]{Jonsson2013a}%
  \BibitemOpen
  \bibfield  {author} {\bibinfo {author} {\bibfnamefont {P.}~\bibnamefont
  {J\"{o}nsson}}, \bibinfo {author} {\bibfnamefont {P.}~\bibnamefont
  {Bengtsson}}, \bibinfo {author} {\bibfnamefont {J.}~\bibnamefont {Ekman}},
  \bibinfo {author} {\bibfnamefont {S.}~\bibnamefont {Gustafsson}}, \bibinfo
  {author} {\bibfnamefont {L.~B.}\ \bibnamefont {Karlsson}}, \bibinfo {author}
  {\bibfnamefont {G.}~\bibnamefont {Gaigalas}}, \bibinfo {author}
  {\bibfnamefont {C.}~\bibnamefont {{Froese Fischer}}}, \bibinfo {author}
  {\bibfnamefont {D.}~\bibnamefont {Kato}}, \bibinfo {author} {\bibfnamefont
  {I.}~\bibnamefont {Murakami}}, \bibinfo {author} {\bibfnamefont {H.~A.}\
  \bibnamefont {Sakaue}}, \bibinfo {author} {\bibfnamefont {H.}~\bibnamefont
  {Hara}}, \bibinfo {author} {\bibfnamefont {T.}~\bibnamefont {Watanabe}},
  \bibinfo {author} {\bibfnamefont {N.}~\bibnamefont {Nakamura}}, \ and\
  \bibinfo {author} {\bibfnamefont {N.}~\bibnamefont {Yamamoto}},\ }\href@noop
  {} {\bibfield  {journal} {\bibinfo  {journal} {Atomic Data and Nuclear Data
  Table}\ ,\ \bibinfo {pages} {in press}} (\bibinfo {year}
  {2013}{\natexlab{b}})}\BibitemShut {NoStop}%
\bibitem [{\citenamefont {{Froese Fischer}}\ \emph {et~al.}(2007)\citenamefont
  {{Froese Fischer}}, \citenamefont {Tachiev}, \citenamefont {Gaigalas},\ and\
  \citenamefont {Godefroid}}]{FroeseFischer2007}%
  \BibitemOpen
  \bibfield  {author} {\bibinfo {author} {\bibfnamefont {C.}~\bibnamefont
  {{Froese Fischer}}}, \bibinfo {author} {\bibfnamefont {G.}~\bibnamefont
  {Tachiev}}, \bibinfo {author} {\bibfnamefont {G.}~\bibnamefont {Gaigalas}}, \
  and\ \bibinfo {author} {\bibfnamefont {M.}~\bibnamefont {Godefroid}},\ }\href
  {\doibase 10.1016/j.cpc.2007.01.006} {\bibfield  {journal} {\bibinfo
  {journal} {Computer Physics Communications}\ }\textbf {\bibinfo {volume}
  {176}},\ \bibinfo {pages} {559} (\bibinfo {year} {2007})}\BibitemShut
  {NoStop}%
\bibitem [{\citenamefont {Cowan}(1981)}]{Cowan1981}%
  \BibitemOpen
  \bibfield  {author} {\bibinfo {author} {\bibfnamefont {R.~D.}\ \bibnamefont
  {Cowan}},\ }\href@noop {} {\emph {\bibinfo {title} {{The theory of atomic
  structure and spectra}}}}\ (\bibinfo  {publisher} {University of California
  Press},\ \bibinfo {year} {1981})\ p.\ \bibinfo {pages} {731}\BibitemShut
  {NoStop}%
\bibitem [{\citenamefont {J\"{o}nsson}\ and\ \citenamefont
  {Gustafsson}(2002)}]{Jonsson2002}%
  \BibitemOpen
  \bibfield  {author} {\bibinfo {author} {\bibfnamefont {P.}~\bibnamefont
  {J\"{o}nsson}}\ and\ \bibinfo {author} {\bibfnamefont {S.}~\bibnamefont
  {Gustafsson}},\ }\href@noop {} {\bibfield  {journal} {\bibinfo  {journal}
  {Computer Physics Communications}\ }\textbf {\bibinfo {volume} {144}},\
  \bibinfo {pages} {188} (\bibinfo {year} {2002})}\BibitemShut {NoStop}%
\bibitem [{\citenamefont {Beck}(2002)}]{Beck2002}%
  \BibitemOpen
  \bibfield  {author} {\bibinfo {author} {\bibfnamefont {D.~R.}\ \bibnamefont
  {Beck}},\ }\href {\doibase 10.1103/PhysRevA.66.034502} {\bibfield  {journal}
  {\bibinfo  {journal} {Phys. Rev. A}\ }\textbf {\bibinfo {volume} {66}},\
  \bibinfo {pages} {034502} (\bibinfo {year} {2002})}\BibitemShut {NoStop}%
\bibitem [{\citenamefont {Zinner}\ \emph {et~al.}(2003)\citenamefont {Zinner},
  \citenamefont {Spoden}, \citenamefont {Kraemer}, \citenamefont {Birkl},\ and\
  \citenamefont {Ertmer}}]{Zinner2003}%
  \BibitemOpen
  \bibfield  {author} {\bibinfo {author} {\bibfnamefont {M.}~\bibnamefont
  {Zinner}}, \bibinfo {author} {\bibfnamefont {P.}~\bibnamefont {Spoden}},
  \bibinfo {author} {\bibfnamefont {T.}~\bibnamefont {Kraemer}}, \bibinfo
  {author} {\bibfnamefont {G.}~\bibnamefont {Birkl}}, \ and\ \bibinfo {author}
  {\bibfnamefont {W.}~\bibnamefont {Ertmer}},\ }\href {\doibase
  10.1103/PhysRevA.67.010501} {\bibfield  {journal} {\bibinfo  {journal}
  {Physical Review A}\ }\textbf {\bibinfo {volume} {67}},\ \bibinfo {pages}
  {010501(R)} (\bibinfo {year} {2003})}\BibitemShut {NoStop}%
\end{thebibliography}%

\end{document}